\documentclass[aps,12pt,superscriptaddress,amsfonts,amssymb,amsmath]{revtex4}

\usepackage{graphicx}
\usepackage{epsfig}
\usepackage{makeidx}
\usepackage{epstopdf}

\begin{document}

\title{Generalized Maxwell equations \\ and charge conservation censorship}

\author{G.\ Modanese \footnote{Email address: giovanni.modanese@unibz.it}}
\affiliation{Free University of Bolzano-Bozen \\ Faculty of Science and Technology \\ I-39100 Bolzano, Italy}

\linespread{0.9}

\begin{abstract}

\bigskip

The Aharonov-Bohm electrodynamics is a generalization of Maxwell theory with reduced gauge invariance. It allows to couple the electromagnetic field to a charge which is not locally conserved, and has an additional degree of freedom, the scalar field $S=\partial_\alpha A^\alpha$, usually interpreted as a longitudinal wave component. By re-formulating the theory in a compact Lagrangian formalism, we are able to eliminate $S$ explicitly from the dynamics and we obtain generalized Maxwell equation with interesting properties: they give $\partial_\mu F^{\mu \nu}$ as the (conserved) sum of the (possibly non-conserved) physical current density $j^\nu$, and a ``secondary'' current density $i^\nu$ which is a non-local function of $j^\nu$. This implies that any non-conservation of $j^\nu$ is effectively ``censored'' by the observable field $F^{\mu \nu}$, and yet it may have real physical consequences. We give examples of stationary solutions which display these properties. Possible applications are to systems where local charge conservation is violated due to anomalies of the ABJ kind or to macroscopic quantum tunnelling with currents which do not satisfy a local continuity equation.


\end{abstract}

\maketitle

\section{Introduction}

Long ago Aharonov and Bohm \cite{AB} proposed ``en passant'' an extension of the electromagnetic Lagrangian, obtained by adding a term $\frac{1}{2} \lambda \left( \partial_\mu A^\mu \right)^2$ to the minimal gauge-invariant Lagrangian which leads to Maxwell equations. A previous attempt in this direction had also been made by Ohmura \cite{Ohm}. This modified electromagnetic theory has been recently re-analysed by Van Vlaenderen and Waser \cite{VV1,VV2}, Hively and Giakos \cite{HG}. In \cite{AB} an Hamiltonian formalism is employed, in \cite{Ohm,VV1} a quaternion formalism, in \cite{VV2,HG} a standard three-dimensional vector formalism. 

In this work we obtain new insights into this extended theory using a standard four-vectors formalism, which is in our opinion more effective, and allows to find some useful explicit expressions missing from the other approaches. We obtain a simple, consistent generalization of the Maxwell equations which is independent from the coupling $\lambda$ and has the important property of charge conservation censorship: although the equations are compatible with sources which do not respect local charge conservation, the measurable field tensor $F_{\mu \nu}$ which solves them is always coupled to a conserved charge.

The results can be relevant for applications to condensed-matter systems with coherent quantum tunnelling, or to more exotic situations in which the occurrence of a chiral anomaly or a spacetime singularity causes a tunnelling process between two vacuum states \cite{che}.

The modified theory of Aharonov-Bohm has an additional degree of freedom, the scalar field $S= \partial_\mu A^\mu$, which cannot be set to zero through a suitable choice of the gauge, as usually done in the standard theory. The field $S$ amounts to the presence, in wave solutions, of an anomalous longitudinal component. Since, however, the source of $S$ is the quantity $\partial_\mu j^\mu$, in fact $S$ is decoupled from matter if charge is locally conserved everywhere. The authors of \cite{HG} discuss the possibility that charge fluctuations generate fluctuations of $S$, or that boundary condition at the interface between vacuum and some media may also affect $S$. Ref.\ \cite{VV2} briefly mentions the possibility that the local conservation of charge is violated in macroscopic tunnelling phenomena. We shall comment on this latter point in our conclusions. 

In Sect.\ \ref{sec2} we show that working in the standard four-vector formalism it is possible to eliminate completely the field $S$ from the field equations, and so obtain generalized, self-consistent Maxwell equations with very interesting properties, namely:
(a) they allow to compute $F_{\mu\nu}$ starting from any physical source $j^\mu$, either locally conserved or not;
(b) if $\partial_\mu j^\mu=0$, the equations reduce to Maxwell equations;
(c) if $\partial_\mu j^\mu \neq 0$, the equations give an additional current $i^\nu$ such that $(j^\nu+i^\nu)$ is conserved and equal to $\partial_\mu F^{\mu\nu}$.

It follows that
any anomalous local non-conservation of the source $j^\mu$ is automatically ``censored'' by the field that it generates, because when we observe the source by measuring $F_{\mu\nu}$ with test particles we can only see the conserved current $(j^\nu+i^\nu)$.
In spite of the censorship, if the physical current $j^\mu$ is not locally conserved, the consequences are real, because the $F_{\mu\nu}$ field will contain a non-maxwellian component whose features will be analysed, in two simple cases, in Sect.\ \ref{sec3}.

Sect.\ \ref{disc} contains our conclusions and a brief discussion of possible applications to Josephson tunnelling in high-T$_c$ superconductors.

\section{Lagrangian and field equations}
\label{sec2}

The Maxwell equations in Heaviside units are written in covariant form as \cite{iz}
\begin{equation}
\begin{array}{*{20}{l}}
{{\partial _\mu }{F^{\mu \nu }} = {j^v}}\\
{{\partial ^\rho }\frac{1}{2}{\varepsilon _{\rho \sigma \mu \nu }}{F^{\mu \nu }} = 0}
\end{array}
\end{equation}
Note that since $F^{\mu \nu }$  is antisymmetric, the field equation for $F^{\mu \nu }$ can only be solved if ${\partial _\nu }{j^\nu } = 0$.
This field equation can be derived from the Lagrangian 
\begin{equation}
L =  - \frac{1}{4}{F_{\mu \nu }}{F^{\mu \nu }} - {j_\mu }{A^\mu } + \frac{1}{2}\kappa \left[ {{{\left( {{\partial _\mu }{A^\mu }} \right)}^2} - {\partial _\mu }{A^\nu }{\partial _\nu }{A^\mu }} \right]
\end{equation}

$L$ is seen as a function of the fundamental dynamical variable  ${A^\mu }(x)$ and is the most general possible relativistic invariant Lagrangian constructed with a four-vector field. The term proportional to $\kappa$  can be written as a four-divergence and therefore gives in fact no contribution to the action  $S = \int {{d^4}xL(x)} $. When $\kappa  = 0$  one speaks of the ``minimal'' Lagrangian. 

In a gauge transformation ${A^\mu } \to {A^\mu } + {\partial ^\mu }\phi $  the variation of $L$ is
\begin{equation}
\Delta L = {j_\mu }{\partial ^\mu }\phi 
\end{equation}
If the current is conserved, this is equivalent to
\begin{equation}
\Delta L = {\partial ^\mu }(\phi {j_\mu }){\rm{\qquad     (if   }}\ \  {\partial _\mu }{j^\mu } = 0)
\end{equation}
and therefore $L$ is gauge-invariant up to a four-divergence. 

Aharonov and Bohm proposed to add a term $\frac{1}{2}\lambda {\left( {{\partial _\mu }{A^\mu }} \right)^2}$  to the minimal $L$. The new addition, not being a four-divergence, changes the field equations, as follows:
\begin{equation}
\begin{array}{l}
{L_{A.B.}} =  - \frac{1}{4}{F_{\mu \nu }}{F^{\mu \nu }} - {j_\mu }{A^\mu } + \frac{1}{2}\lambda {\left( {{\partial _\mu }{A^\mu }} \right)^2} \to \\
 \to {\partial _\mu }{F^{\mu \nu }} = {j^\nu } + \lambda {\partial ^\nu }\left( {{\partial _\alpha }{A^\alpha }} \right)
\end{array}
\label{field1}
\end{equation}
Under a gauge transformation, the variation of the Aharonov-Bohm Lagrangian is
\begin{equation}
\Delta {L_{A.B.}} = {j_\mu }({\partial ^\mu }\phi ) + \frac{1}{2}\lambda [{({\partial ^\alpha }{\partial _\alpha }\phi )^2} + 2({\partial ^\alpha }{A_\alpha })({\partial ^\alpha }{\partial _\alpha }\phi )]
\end{equation}
Thus the Lagrangian is not gauge-invariant anymore. It is only invariant (if  ${\partial _\mu }{j^\mu } = 0$) under reduced gauge transformations, such that  ${\partial ^\alpha }{\partial _\alpha }\phi  = 0$. 

In the Aharonov-Bohm theory we cannot impose the familiar Lorenz gauge, choosing a new four-potential  $A{'^\mu }$ such that ${\partial _\mu }A{'^\mu } = 0$. Therefore we must regard the quantity $S = {\partial _\alpha }{A^\alpha }$  as a non-trivial dynamical variable. 
Now take the derivative ${\partial _\nu }$  of the field eq.\ (\ref{field1}) and remember that ${F^{\mu \nu }}$   is antisymmetric. We obtain
\begin{equation}
{\partial _\nu }{j^\nu } =  - \lambda {\partial _\nu }{\partial ^\nu }({\partial ^\alpha }{A_\alpha }) =  - \lambda {\partial ^2}S
\label{field2}
\end{equation}
where ${\partial ^2}$  denotes the D'Alembert operator  ${\partial ^2} = {\partial _\alpha }{\partial ^\alpha }$. 
From (\ref{field2}) we obtain 
\begin{equation}
S = {\partial _\alpha }{A^\alpha } =  - \frac{1}{\lambda }{\partial ^{ - 2}}\left( {{\partial _\nu }{j^\nu }} \right)
\label{field3}
\end{equation}
The well-known operator ${\partial ^{ - 2}}$  is linear and non-local, as can be seen passing to four-momentum space, where it is represented by ${k^{ - 2}}$ . Eq.\ (\ref{field3}) allows us to write an expression for the variable $S$, essentially integrating over the ``source'' ${\partial _\nu }{j^\nu }$:
\begin{equation}
S(x) =  - \frac{1}{\lambda }\int {{d^4}k{e^{ - ikx}}\frac{{{k_\nu }{{\tilde j}^\nu }(k)}}{{{k^2}}}} 
\end{equation}

Now go back to (\ref{field2}), rename the summation index ($\nu \to \beta$), take again the derivative  ${\partial ^\nu }$ and multiply by $\lambda$. We obtain
\begin{equation}
\lambda {\partial ^\nu }\left( {{\partial _\alpha }{A^\alpha }} \right) =  - {\partial ^\nu }{\partial ^{ - 2}}\left( {{\partial _\beta }{j^\beta }} \right)
\end{equation}
Plug this into (\ref{field1}) and re-write the inhomogeneous generalized Maxwell equations as follows (the homogeneous equations remain the same):

{\bf Version 1}: Equations for the field strength ${F_{\mu \nu }}$, the most simple and useful:
\begin{equation}
\left\{ \begin{array}{l}
{\partial _\mu }{F^{\mu \nu }} = {j^\nu } + {i^\nu }  \\
{i^\nu } =  - {\partial ^\nu }{\partial ^{ - 2}}\left( {{\partial _\beta }{j^\beta }} \right)
\end{array} \right.
\label{mme1}
\end{equation}

{\bf Version 2}: Equations for the four-potential  ${A_\mu }$, formally more complex:
\begin{equation}
\left\{ \begin{array}{l}
{\partial _\mu }({\partial ^\mu }{A^\nu } - {\partial ^\nu }{A^\mu }) = {j^\nu } + {i^\nu }\\
{i^\nu } =  - {\partial ^\nu }{\partial ^{ - 2}}\left( {{\partial _\beta }{j^\beta }} \right)\\
{\partial _\alpha }{A^\alpha } =  - \frac{1}{\lambda }{\partial ^{ - 2}}\left( {{\partial _\nu }{j^\nu }} \right)
\end{array} \right.
\label{mme2}
\end{equation}

Note that eq.\ (\ref{mme2}-c) is technically an awkward gauge condition, ``obtainable'' with the residual gauge freedom. A quantum field theory of the four-potential in this gauge looks difficult! Yet we believe the classical field equations of Version 1 are interesting enough.
Note that although these are derived from the Aharonov-Bohm Lagrangian ${L_{A.B.}}$, the parameter $\lambda$ has disappeared. If the current ${j^\mu }$  is conserved, then the usual Maxwell equations are recovered. The new current component ${i^\nu }$  which now contributes, together with  ${j^\nu }$, to generate the field  ${F^{\mu \nu }}$, is such that the total current $\left( {{j^\nu } + {i^\nu }} \right)$  is always conserved, as can be checked in two ways: (1) by taking the derivative ${\partial _\nu }$  in the first equation of (\ref{mme1}); (2) by taking the derivative ${\partial _\nu }$  in eq.\ (\ref{mme1}-b), which yields consistently
\begin{equation}
{\partial _\nu }{i^\nu } =  - {\partial _\nu }{\partial ^\nu }{\partial ^{ - 2}}\left( {{\partial _\beta }{j^\beta }} \right) =  - {\partial _\beta }{j^\beta } \Rightarrow {\partial _\nu }\left( {{j^\nu } + {i^\nu }} \right) = 0
\end{equation}

Summarizing, we can say that the input of the generalized electrodynamic equations (\ref{mme1}) is a four-current  ${j^\nu }$ which is not necessarily conserved (computed, for instance, from an ``anomalous'' microscopic model, as discussed in the following); but the output is an electromagnetic field tensor ${F^{\mu \nu }}$  which has the usual properties, including that of being generated by a conserved current, namely  $\left( {{j^\nu } + {i^\nu }} \right)$. It follows the important property that at the macroscopic level the current is always conserved, as far as it is possible to measure it through the field it generates. Since eq. (\ref{mme1}-a) is linear, the field ${F^{\mu \nu }}$  is the sum of the fields generated by the currents ${j^\nu }$  and  ${i^\nu }$. In general, the difference between the two currents is that even if the ``primary'' current $j^\nu$  is confined in a certain region of spacetime, the ``secundary'' current ${i^\nu }$  is not, because of the non-local expression which relates it to  ${j^\nu }$. A further discussion of this point is given in Sect.\ \ref{disc}.

\section{Stationary solutions with planar and dipole sources}
\label{sec3}

Let us find the solutions of the generalized Maxwell equations (\ref{mme1}) for a simple planar source in stationary flow regime $\frac{\partial \rho}{ \partial t }=0$ (Fig.\ 1). We assume that inside the source the three-divergence $\nabla  \cdot {\bf{j}}$ is not zero, i.e., charge is being emitted (possibly by tunnelling from another region where it is absorbed; compare Fig.\ 2). The source is a thin layer of thickness $d$ on the $x$-$y$ plane. Suppose that inside the layer we have $\frac{\partial j_z}{ \partial z } > 0$ (Fig.\ 1, (a)). The current density ${\bf j} = \eta {\hat {\bf z}}$ will generate, according to eq.\ (\ref{mme1}-a), a magnetostatic field in the space region with $z>0$. There will also be, however, an additional magnetostatic field generated by the secondary current ${\bf i}$.

\begin{figure}
\begin{center}
\includegraphics[width=10cm,height=7cm]{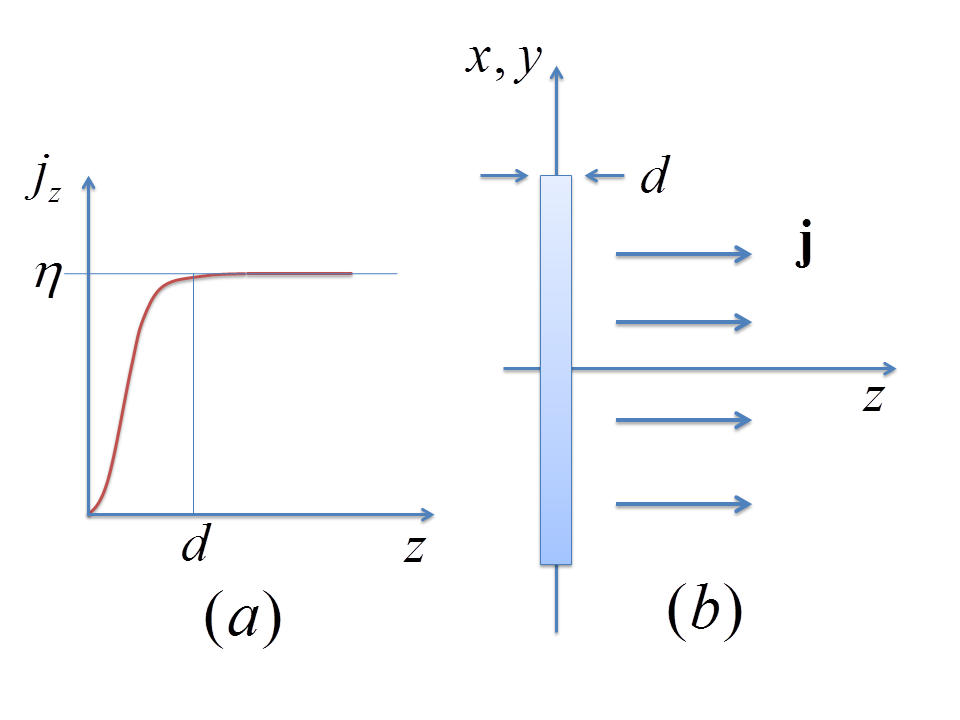}
\caption{Physical current density ${\bf{j}} = \eta {\bf{\hat z}}$ for a plane which emits a stationary flow of charge, thus violating the continuity condition ${\partial _\alpha }{j^\alpha } = 0 \to \nabla  \cdot {\bf{j}} = 0$. The secondary current density (eq.\ (\ref{mme1}-b)) is ${\bf{i}} = \pm \frac{1}{2}\eta {\bf{\hat z}}$ (+ for $z<0$, -- for $z>0$), so the violation is censored by the total current ${\bf{j}} + {\bf{i}}$. Compare Fig.\ 2, where the secondary currents are explicitly drawn.} 
\label{f01}
\end{center}  
\end{figure}

Let us compute ${\bf i}$. 
Start from eq.\ (\ref{mme1}-b) and rewrite it as ${i^\nu } = {\partial ^{ - 2}}{\partial ^\nu }({\partial _\alpha }{j^\alpha })$. In this case ${\partial _\alpha }{j^\alpha }$ does not depend on time. Then $i^0=0$ and for the spatial component $i^k$ we can write
\begin{equation}
{i^k} = {\partial ^k}{\partial ^{ - 2}}({\partial _\alpha }{j^\alpha })
\end{equation}

But ${\partial ^{ - 2}}({\partial _\alpha }{j^\alpha })=\Delta^{-1}({\partial _\alpha }{j^\alpha })$ is, in Heaviside units, like $\frac{1}{2}$ the electric potential of a charge distribution with density ${\partial _\alpha }{j^\alpha }$; and $-{\partial ^k}{\partial ^{ - 2}}({\partial _\alpha }{j^\alpha })$ is like the electric field of this distribution. These are mathematical properties familiar from the Maxwell equations, which we may apply here, although in a different context.

The source of this ``equivalent electric field'', which coincides mathematically with the secondary current ${\bf i}$, is a planar charge distribution with surface density
\begin{equation}
\sigma=\int dz \, \nabla  \cdot {\bf{j}} = \int dz \, \frac{\partial j_z}{ \partial z } = \eta
\end{equation}

Therefore the secondary current density ${\bf i}$ enters into the $x$-$y$ plane both from the right and the left (like the electric field of a negatively charged plane):
\begin{equation}
\begin{array}{l}
{\bf{i}}({\bf{x}}) = -\frac{1}{2} \eta {\bf{\hat z}}{\rm{\qquad   for }}\qquad z \ge 0\\
{\bf{i}}({\bf{x}}) = \frac{1}{2}  \eta {\bf{\hat z}}{\rm{\qquad   for }}\qquad z < 0
\end{array}
\end{equation}

It follows that the total current density is continuum:
\begin{equation}
{\bf{j}} + {\bf{i}} = \frac{1}{2}\eta {\bf{\hat z}}
\end{equation}
The emission of charge from the plane is censored, since an observer who can only measure the electromagnetic field will see a continuum flux of charge going from $z=-\infty$ to $z=+\infty$. For the case of a double plane, see Fig.\ 2.

\begin{figure}
\begin{center}
\includegraphics[width=10cm,height=7cm]{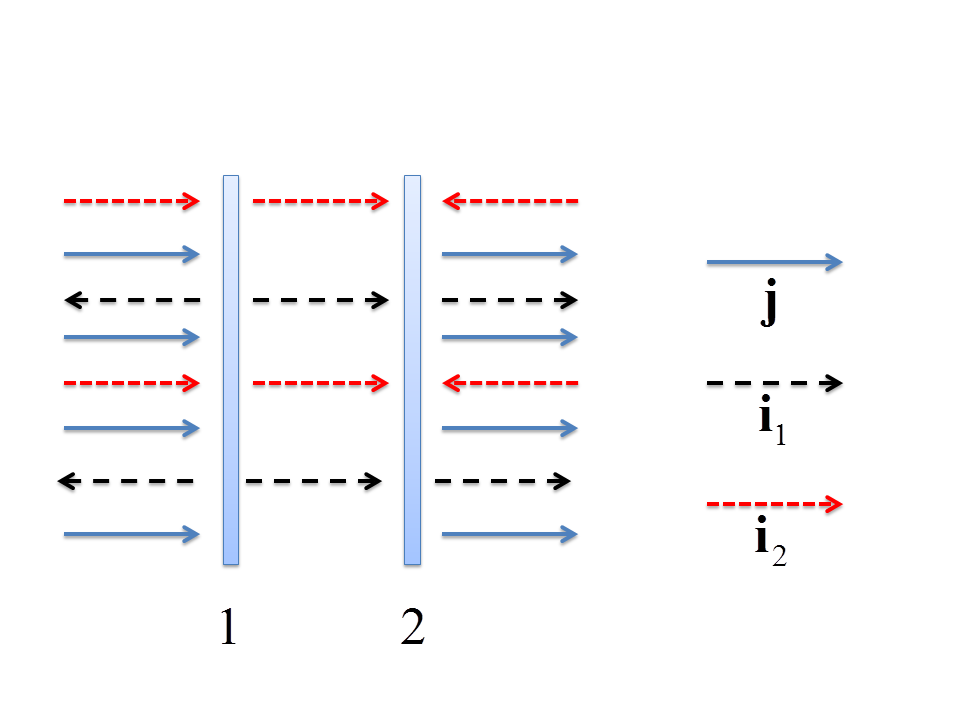}
\caption{Parallel planes which absorb and re-emit current, with a tunnelling process which violates the continuity equation (a process whose wave function does not satisfy the Schroedinger or Ginzburg-Landau equations). Each arrow represents schematically one unit of current density. The violation is completely censored by the total current ${\bf{j}} + {\bf{i}_1}+{\bf{i}_2}$, which amounts to ``four units'' everywhere. ${\bf{j}}$: physical, or ``primary'', current density. ${\bf{i}_1}$: secondary current density emitted by Plane 1 according to eq.\ (\ref{mme1}-b). ${\bf{i}_2}$: secondary current density emitted by Plane 2.} 
\label{f02}
\end{center}  
\end{figure}

Next consider a stationary dipolar source made of a pointlike charge emitter placed near a pointlike absorber, i.e.\
\begin{equation}
{\partial _\alpha }{j^\alpha } = \nabla  \cdot {\bf{j}} = \mu \left[ {{\delta ^3}\left( {{\bf{x}} - {\bf{a}}} \right) - {\delta ^3}\left( {{\bf{x}} + {\bf{a}}} \right)} \right]
\end{equation}
Let us just compute the secondary current, in order to check that it is not localized:
\begin{equation}
{i^k} = {\partial ^k}{\Delta ^{ - 1}}\mu \left[ {{\delta ^3}\left( {{\bf{x}} - {\bf{a}}} \right) - {\delta ^3}\left( {{\bf{x}} + {\bf{a}}} \right)} \right] = \frac{1}{2}\mu {\partial ^k}\left( {\frac{1}{{\left| {{\bf{x}} - {\bf{a}}} \right|}} - \frac{1}{{\left| {{\bf{x}} + {\bf{a}}} \right|}}} \right)
\label{i-dip}
\end{equation}
Therefore $i^k$ is like the electric field of a dipole. The magnetostatic field that it generates can be computed from (\ref{mme1}-a), for instance through the Biot-Savart formula.

\section{Discussion and conclusions}
\label{disc}

We have seen that the generalized Aharonov-Bohm electrodynamics, here re-formulated and solved for the scalar field $S$, is applicable also to situations where charge is not locally conserved. In these situations the electromagnetic field tensor $F_{\mu \nu}$ hides the non-conservation, because its source contains also an additional, ``secondary'' current term which compensates for the non-conserved part of the physical current. This mechanism is particularly clear in the example of the double plane with tunnelling (Fig.\ 2).

One might thus have the feeling that in the end the theory is not telling us anything new, compared to Maxwell theory. This is not the case, however, since the field generated by the secondary current $i^\nu$ is not maxwellian, as a consequence of the non-local character of $i^\nu$. Such non-locality descends in general from eq.\ (\ref{mme1}-b) and is clearly seen in the example of the static dipole, eq.\ (\ref{i-dip}). (An oscillating dipole solution will be presented in a separate paper.)

What do we mean exactly by ``non-maxwellian'' field? Given any antisymmetric tensor $F^{\mu \nu}$, according to (\ref{mme1}-a) the derivative $\partial_\mu F^{\mu \nu}$ yields a conserved source $(j^\nu+i^\nu)$ which generates it. If we substitute this entire source into the Maxwell equations and regard it as a physical source, then we obtain the same $F^{\mu \nu}$ as a solution of the Maxwell equations, and so by definition ``maxwellian''. It may happen, however, that $(j^\nu+i^\nu)$ is manifestly not physical, for instance because it extends to spatial infinity while the material systems we are considering are limited in space. In this case we say that $F^{\mu \nu}$ is not maxwellian.

More specifically, we may wonder if there exists a solution $F^{\mu \nu}_L$ which represents a wave in vacuum ($j^\nu=0$) with a longitudinal component. From (\ref{mme1}-a) we obtain $i^\nu$: $i^\nu_L=\partial_\mu F^{\mu \nu}_L$. Eq.\ (\ref{mme1}-b) relates $i^\nu$ to the physical current density $j^\nu$. The question then becomes: does a suitable physical current $j^\nu$ exist, located inside a bounded region, able to generate in the vacuum space near it a current $i^\nu_L$? A detailed answer to this question will require to analyse the invertibility of eq.\ (\ref{mme1}-b). The answer might be affirmative thanks to the non-locality of the operator $\partial^{-2}$. Compare also the functional approach in \cite{cqg}.

Finally, let us briefly discuss the possible concrete existence of physical currents which are not locally conserved. The ABJ anomalies represent, as mentioned, such a possibility. We recall that ABJ anomalies have also been observed in condensed matter \cite{PRX}, but as anomalies in momentum space, and therefore probably not relevant to our case.

The possible non-conservation of current in macroscopic quantum tunnelling is discussed in \cite{VIG}, with reference to Josephson junctions in high-T$_c$ superconductors. In short, it is conjectured that due to the very small correlation length $\xi$, the stationary continuity condition $\nabla \cdot {\bf j}=0 \to \rho v=const.$ may be violated in wide inter-grain junctions where $\rho$ decreases by many magnitude orders \cite{Hilge,Andoh} while $v$ cannot increase in inverse proportion.
It is known that the solutions of the Ginzburg-Landau equation satisfy a continuity condition for the current. The Ginzburg-Landau equation, however, should be only regarded as a phenomenological local approximation. The microscopic theory of high-T$_c$ superconductors is still unsettled. In complex condensed matter systems, macroscopic wave functions of charge carriers obey a constrained equation and have therefore in general a non-local conserved current \cite{Schw}. This might be related to the puzzle of the superluminal tunnelling times \cite{Landauer,Cardone}.

In conclusion, it cannot be ruled out that in certain situations the macroscopic tunnelling of the wave function is associated to non-local conservation of the current, thus requiring the generalized Aharonov-Bohm theory, and possibly generating non-maxwellian fields in the sense discussed above.

Furthermore, the possible existence of non-conserved currents associated to tunnelling processes is obviously relevant also for other interactions, besides the electromagnetic interaction. One can consider, for instance, non-conserved hadronic currents. The generalized electrodynamics presented in this paper can be applied to such currents and tells us that they could be partially concealed from electromagnetic measurements, but still be real.


\begin{thebibliography}{100}

\bibitem{AB}
Y. Aharonov and D. Bohm, Phys. Rev. 130 (1963) 1625.

\bibitem{Ohm}
T. Ohmura, Prog. Theor. Phys. 16 (1956) 684-685.

\bibitem{VV1}
K.J. Van Vlaenderen and A. Waser, Hadronic J. 24 (2001) 609-628. 

\bibitem{VV2}
K.J. Van Vlaenderen, ``A generalisation of classical electrodynamics for the prediction of scalar field effects'', arXiv: physics/0305098.

\bibitem{HG}
L.M. Hively and G.C. Giakos, Int. J. Signal and Imaging Systems Engineering 5 (2012) 3-10.

\bibitem{che}
T.-P. Cheng and L.-F. Li, {\it Gauge theory of elementary particle physics}, Clarendon, Oxford, 2000.

\bibitem{iz}
C. Itzykson and J.-B. Zuber, {\it Quantum field theory}, Dover, New York, 2005

\bibitem{cqg}
G. Modanese, Class. Quantum Gravity 24 (2007) 1899.

\bibitem{PRX}
S.A. Parameswaran et al., Phys. Rev. X 4 (2014) 031035.

\bibitem{VIG}
G. Modanese, ``Covariant formulation of Aharonov-Bohm electrodynamics and its application to coherent tunnelling'', in {\it Unified Field Mechanics II}, Proceedings of the 10th Symposium in Honour of J.-P. Vigier, Portonovo, 2016. 

\bibitem{Hilge}
H. Hilgenkamp and J. Mannhart, Rev. Mod. Phys. 74 (2002) 485.

\bibitem{Andoh}
H. Andoh et al., Physica C 339 (2000) 237-244.

\bibitem{Schw}
M. Schwartz, Y. Navot, Physica A 245 (1997) 517-522.

\bibitem{Landauer}
R. Landauer and T. Martin, Rev. Mod. Phys. 66 (1994) 217

\bibitem{Cardone}
F. Cardone and N. Mignani, {\it Deformed Spacetime: Geometrizing Interactions in Four and Five Dimensions}, Springer, Dordrecht, 2007.


\end{thebibliography}
\end{document}